\documentclass[aps,epsfig]{revtex4}

\usepackage[centertags]{amsmath}
\usepackage{amssymb}
\usepackage{amsthm}
\usepackage{graphicx,subfigure}
\usepackage{epsfig}

\begin{document}

\title{Classical Open Systems coupled to Nonlinear Baths: Noise Spectrum and Dynamical Correlations}

\author{Chitrak Bhadra, Ion Santra and Dhruba Banerjee}


\affiliation{Department of Physics, Jadavpur University, Kolkata, India}

\begin{abstract}
\noindent Open system dynamics in a classical setting is microscopically governed by the structure of the thermal environment which influences the dynamics of the probe particle (free or in an external potential). Nonlinear baths have recently been shown to impart interesting nonequilibrium correlations in the dissipative dynamics affecting Generalised Langevin Equations and the Fluctuation Dissipation Relations. In the following work, we investigate some aspects of nonlinear baths with rigour relying on perturbative expansions to deal with nonlinearities. Firstly, the question of noise spectrum emerging from such nonlinearities are addressed and the Markovian limit is explored; super-Ohmic corrections to the linear Ohmic spectrum is deduced. Velocity correlations of a probe system under such approximations are studied in detail. In a second part to the paper, a quenched initial thermal bath is modelled via nonlinearities and Louivillean evolution is applied to evaluate subsequent correlations out of equilibrium. In all these problems, weak system-bath coupling is assumed and quartic baths are used for specific calculations.
\end{abstract}

\maketitle

 \section{Introduction}


The study of dissipative systems, classical and quantum, in the presence of random forces (noise) has been in focus for more than a century. At the classical level, the mathematical paradigm is based on the seminal work of Paul Langevin, a stochastic differential equation formulation later extended by Ornstein and Uhlenbeck \cite{chandra,lang,orn}. It describes the motion of a system under an external potential with two important ingredients: dissipation of energy over time and random noise, both arising due to interaction with an environment (we will call it a bath throughout the paper). In the limit of Gaussian white noise (Markovian limit) and an initial equilibrium thermal distribution of the bath modes, the Langevin equation describes Brownian Motion in the continuum and the diffusion co-efficient emerges as a natural parametrisation of the motion. On the other hand, Generalised Langevin Equations (GLE) have been investigated over the decades which explain effects of nonlinear damping and non-Markovian, non-Gaussian coloured noise \cite{zwa1}. The phenomenology explained by Langevin equations covers a vast range of stochastic dynamical phenomena in nature and dictates experiments at small length scales. The GLE is laid out below in its most generic form for a probe of mass $M$, in an external potential $V(X)$, a dissipation kernel $\gamma (t)$ and state-dependent noise $\Gamma(t)$.

\begin{eqnarray}
M\ddot{X} + M\Omega ^2 X + M \int_{0}^{t}\gamma (t-t^{'})\dot{X}(t')dt^{'} &=&\Sigma (X,t) \Gamma(t)
\end{eqnarray}\\

\noindent A GLE describes the coarse-grained effect of the bath on the system itself and a different outlook was presented through a system-bath Hamiltonian approach by Zwanzig,Mori and others \cite{zwa1,mori,zwa2,kawa} to account for a microsopic description of classical GLEs. Intuitively, the picture is simple: since dissipation and noise arise out of the interaction between system and the environment in which its dynamics occurs, an explicit partioning of the Hamiltonian is done. This contains the system, bath and interaction Hamiltonian systematically and subsequent equation of motion under a probabilistic interpretation of initial conditions for the bath lead to a coarse-grained dynamical equation representing a GLE \cite{weiss}. Gaining a handle on the microscopic nature of the problem, nonlinearities and coloured noise could be incorporated into the description of a much extended domain of open systems and most importantly, quantisation of the dynamics can be carried out \cite{feyn,heinz}. Most importantly, an explicit derivation of Fluctuation-Dissipation Relations (FDR) \cite{kubo}, which relate the damping kernel with the noise correlations, emerge naturally. The FDR at equilibrium can be thought of as the consistency check for an open system dynamics and its violations and retrieval of Fluctuation relations have spanned a mammoth literature when it comes to out-of-equilibrium Physics too \cite{vul2}. 

\begin{eqnarray}
\langle \Gamma(t)\Gamma (t')\rangle \propto  \gamma (t-t') 
\end{eqnarray}

\noindent In a series of papers \cite{bhadra1,bhadra2} we have addressed the classical problem of a general system coupled linearly to a bath of nonlinear oscillators. Nonlinear reservoirs have not come into direct focus from the microscopic viewpoint, where one employs Zwanzig's approach for forming the generalized Langevin equation by eliminating the bath degrees of freedom. Here we extend that program to study the effect of nonlinearities in the bath oscillators through perturbation theory. The takeaway from those investigations was the fact that the Fluctuation Dissipation Relation undergoes some modification order by order and this modification involves nonlinear powers of $k_BT$ in addition to the usual FDR of the second kind (for harmonic baths) that we retrieve at the zeroeth order. The interesting aspect that transpires from the structure of the FDR is that in \textit{weak system-bath coupling regime} the damping term gets renormalized through the coupling constants which become temperature dependent as a result of the bath nonlinearities. Thus we understand that the correct damping term to start with in the Generalized Langevin Equation is the renormalized damping instead of the ``bare" damping one usually encounters for harmonic baths. On the other hand, a \textit{strong system-bath coupling} leads to new nonlinear dissipation functions in the GLE itself and complicated nonlinear, nonlinear correlations appear signalling the emergence of a purely nonequilibrium paradigm for the evolution of the probe system.\\

\noindent However, as in so many situations in physical systems, to understand the dynamics of a probe particle in a background of nonlinear thermal environment, the weak coupling limit produces much more meaningful result. This leads us to our sojourn into evaluating the effect on specific systems in the following pages. We take up completely two different persepectives and present the results in the light of what have already been uncovered about nonlinear baths. Firstly, the FDR of the first kind, ie, the velocity-velocity correlations of the probe particle (under the action of a specific external potential) is an important framework of determining the coarse-grained effects of a bath \cite{risken,kamp}. For e.g., the velocity correlations can easily lead to a statement about the equipartition theorem for a particle in equilibrium. The spatial correlations on the other hand lead to the diffusion coefficient of a dissipative probe particle. Notably, the most important ingredient in calculating these quantities is the the noise correlations and the bath spectrum. This relation serves as a two-way street connecting the microscopics and macroscopics of open systems. In fact, depending on the nature of Markovian/Non-markovian feature of the correlations, the bath spectrum can be rigorously determined. Generally, when attacking problems of specific probe particles, a Markovian, memoryless noise is assumed to fecilitate calculation. This is the prototypical white or Gaussian noise. The system-bath formalism naturally gives a non-Markovian formula for the noise-correlations and thus extending the space of possible Langevin equations to include non-Gaussian noise. The Ohmic approximation of the noise spectrum to the noise-spectrum naturally recover the above mentioned Markovian as well. And for nonlinear baths, such a stage is already set.\\

\noindent In a second part to this paper, we look at the interesting possibility of a time-dependent perturbation on a thermal bath (again, in a weak system-bath coupling regime). The description of open system dynamics is founded on the canonical ensemble at thermal equilirbium for the bath modes which are generically considered quadratic in nature. The equilibrium initial condition is protected through the FDR and can be easily seen through the velocity correlations of a probe particle. All excess correlations decay at large time scales. Our forays into nonlinear baths help us set up an ideal situation for non-stationary baths: at the moment of initialisation of the open system dynamics, a sudden quench occurs in the bath spectrum which we model via a nonlinear perturbation on the quadratic modes. This immediately leads to the time evolution of the probability distribution governing the dynamics through the introduction of the Louiville operator \cite{liu}. Time-dependent correlations emerge as a subset from the stationary case and a perturbative analysis is provided herein laying down the most important terms that regulate the dynamics of the system. Strictly, this example is that of a fully non-equilibrium problem involving time-dependent baths and may serve as a basis for many other situations in physical systems out of equilibrium. A summary and Appendix round off the paper in the last two sections. 

\section {Noise spectrum and Velocity Correlations}

\noindent Macroscopic dynamical correlations of an open system in a thermal environment is intrinsically related to the microscpic structure of the bath correlations and spectrum. We lay down this path step by step in a nonlinear bath paradigm and retrieve the limit of Markovian dynamics, the fundamental setting of any stochastic process.\\

\subsection{From noise correlations to velocity correlations: a simple derivation}

\noindent We begin by some basic reviews of derivation of velocity-correlations of the probe system, the importance of the noise-correlations through its connection to the FDR of the second kind and the structure of a realistic noise spectrum \cite{risken}. The standard Gaussian white-noise influenced Langevin equation for a free particle is given by the equation

\begin{eqnarray}
M\dot{V} + M\gamma V &=&\Gamma(t)
\label{eq100}
\end{eqnarray}

\noindent The averages of the stochastic forces $\Gamma(t)$ are well-known:

\begin{eqnarray}
\langle \Gamma(t)\rangle &=& 0 \nonumber\\
\langle \Gamma(t)\Gamma (t')\rangle &=& \frac {2 K_{B}T}{M} \delta (t-t') 
\label{eq101}
\end{eqnarray}

\noindent where, the spectral density of $S(\omega)$ of the noise forces is evaluated through the Weiner-Kinchin theroem \cite{kamp} and is found to be frequency independent,

\begin{eqnarray}
S(\omega)&=& 2 \int_{-\infty} ^{+\infty} \exp (- i \omega \tau) \frac{K_{B}T}{M} \delta (\tau) d\tau \nonumber\\
&&= \frac{4 K_{B}T}{M}
\label{eq102}
\end{eqnarray}

\noindent To solve the inhomogeneous differential equation \ref{eq100}, we impose a sharp initial condition on the velocity of the probe particle $V_{0}$. This is also commensurate with the fact that the system and the bath are considered uncoupled at initial times. We arrive at,
\begin{eqnarray}
V(t) &=& V_{0} \exp (-\gamma t) + \int _{0} ^{t} \exp (-\gamma (t-t')) \Gamma (t') dt'
\label{eq103}
\end{eqnarray}

\noindent And, immediately the correlation function for the free probe particle is determined.

\begin{eqnarray}
\langle V(t)V(t')\rangle &=& V_{0} ^2 \exp (-\gamma (t_{1} + t_{2}))\nonumber\\
&+&  \int _{0} ^{t_{1} } dt'_{1}\int _{0} ^{t_{2}} dt'_{2} \exp (-\gamma (t_{1}+t_{2}-t'_{1}-t'_{2})) \langle \Gamma(t'_{1})\Gamma (t'_{2})\rangle \nonumber\\
\label{eq103}
\end{eqnarray}

\noindent Evaluating the double integral, first the integration is carried over $t'_{2}$ and then from zero to any of the time variables provided it is the minimum of the two. Using the condition of $\delta$-correlated noise, this leads to an exact expression for the velocity correlation.

\begin{eqnarray}
\langle V(t)V(t')\rangle &=& V_{0} ^2 \exp (-\gamma (t_{1} + t_{2})) + \frac{K_{B}T}{M} \Big (\exp (-\gamma (t_{1} + t_{2}) + \exp (-\gamma (t_{1} - t_{2}) \Big)\nonumber\\
\label{eq104}
\end{eqnarray}

\noindent For large times the familiar result for the FDR of the first kind is formulated.

\begin{eqnarray}
\langle V(t)V(t')\rangle &=& \frac{K_{B}T}{M} \exp \left[-\gamma (t_{1} - t_{2})\right]
\label{eq105}
\end{eqnarray}

\subsection{System-Bath perspective and spectral distribution of the baths}

\noindent The above derivation highlights the powerful method of the Langevin equation and the ease of calculating associated dynamical correlations. However, noise in many dynamical, chemical and biological systems violate $\delta$-correlated conditions \cite{jung}. These are non-Markovian models where the noise forces are correlated in time (spatial correlations are excluded in our discussion) and have a strong bearing on the noise spectrum of the bath and a microscopic viewpoint in this context is necessiated. With the system-bath formalism already at our disposal, we embark on the implications of such a scenario and develop a comprehensive understanding of nonlinear bath spectrum in the weak coupling limit. But first, the simple quadratic bath is explored which does not automatically mean the existence of white noise; physical approximations are made to regularise such baths (leading to a white or non-Gaussian noise forms) and pave the way for studying perturbatively added nonlinearities.\\

\noindent The coarse-grained GLE for a generic system coupled to a quadratic bath takes the form
\begin{eqnarray}
M\ddot{X} + M \int_{0}^{t}\gamma (t-t^{'})\dot{X}(t')dt^{'} &=&\Gamma(t)
\label{eq106}
\end{eqnarray}

\noindent where the time non-local damping kernel is given by the standard form

\begin{eqnarray}
\gamma (t-t^{'})&=& \lambda^{2} \Theta (t-t') \sum_{\mu}\frac{C^{2}_\mu}{m_{\mu} \omega_{\mu}^2} \cos[\omega_{\mu}(t-t')]
\label{eq107}
\end{eqnarray}

\noindent The theta-function suffices to address the fact that the dynamics is evaluated for all $t'$ upto time t. Now, the Fourier Transform of the dissipation kernel is given by \cite{weiss}

\begin{eqnarray}
\tilde {\gamma}(\omega) &=& \lim _{\eta \to 0^+} \frac{-i \omega}{M}\sum_{\mu}\frac{C^{2}_{\mu}}{m_{\mu} \omega_{\mu}^2} \frac {1}{\omega _{\mu} ^2 - \omega ^2 + i\eta \omega}
\label{eq108}
\end{eqnarray}

\noindent The spectral density of the bath (a physically relevant quantity) is defined as 

\begin{eqnarray}
S(\omega) &=& \frac{\pi}{2} \sum_{\mu}\frac{C^{2}_{\mu}}{m_{\mu}\omega_{\mu}} \delta (\omega - \omega _{\mu})
\label{eq109}
\end{eqnarray}

\noindent making the transition to a realistic continuum of bath modes lead to the final compact form for the Fourier Transformed damping kernel 

\begin{eqnarray}
\tilde {\gamma}(\omega) &=& \lim _{\eta \to 0^+} \frac{-i \omega}{M} \frac{2}{\pi} \int _{0}^{\infty} d\omega ' \frac{S(\omega ')}{\omega '} \frac {1}{\omega '^2 - \omega ^2 + i\eta \omega}
\label{eq110}
\end{eqnarray}\\

\noindent immediately gives $\gamma (\omega)$ a smooth real part, which can be derived by calculating the imaginary part of the $\omega$ integral, employing the residue theorem for the contour around the pole at positive $\sqrt{\omega ^2 - i\eta \omega}$ and letting the limit $\eta $ go to zero (see Appendix).

\begin{eqnarray}
\tilde{\gamma}' (\omega) &=& \frac{S(\omega)}{M \omega} 
\label{eq111}
\end{eqnarray}

\noindent Rewriting the above equation explicitly, we get

\begin{eqnarray}
S(\omega) &=& M \omega \int _{0} ^{\infty} dt \gamma (t) \cos \omega t
\label{eq112}
\end{eqnarray}

\noindent This relation lays the platform for calculating the noise spectrum directly from the time-real damping kernel which defined the FDR via equivalence to the noise correlations. For frequency-dependent damping kernels, the Laplace transform of $\gamma (t)$ is also straightforward and useful.

\begin{eqnarray}
\tilde {\gamma} (s) &=& \tilde {\gamma}(\omega = i s) \nonumber\\
\tilde {\gamma} (s) &=& \frac{s}{M} \frac{2}{\pi} \int _{0}^{\infty} d\omega ' \frac{S(\omega ')}{\omega '} \frac {1}{\omega '^2 + s^2}
\label{eq113}
\end{eqnarray}

\noindent The above formula can be used via an Inverse Laplace Transform to give 

\begin{eqnarray}
\gamma (t) &=& \Theta (t) \frac{2}{M \pi} \int _{0}^{\infty} d\omega  \frac{S(\omega)}{\omega} \cos \omega t
\label{eq114}
\end{eqnarray}

\noindent The macroscopic dissipation function thus is completely determined through the integrated-out effects of the microscopic degrees of freedom of the underlying bath structure. For all practical purposes, the dynamics of the probe system is established by the parameters M, external potential $V(X)$ and the noise spectrum $S(\omega)$ of the bath.\\

\noindent The above analysis reflects the two-way route to evaluating the effect of an environment on open system dynamics. On one hand, if a certain bath structure is established (naturally occuring or engineered through artificial means), we have a handle on predicting the dissipation function in the weak system-bath coupling limit and correspondingly the short and long-term dynamical correlations. On the other hand, if we start from a microscopic picture and go backwards to build such a noise-spectrum or understand the microscopic structure of the bath, the above formalism suffices. This is an important observation, as even in a highly complicated scenario of the underlying Hamiltonian structure, Langevin dynamics is established as an emergent macroscpic phenomenon with a small number of variables governing the dynamics.\\

\subsubsection{ Gaussian white noise: Ohmic approximation}

\noindent As has been mentioned previously, the white noise is defined by the memoryless correlation function

\begin{eqnarray}
\langle \Gamma(t)\Gamma (t')\rangle &=& A \delta (t-t') 
\label{eq115}
\end{eqnarray}

\noindent with the damping kernel taking a $\delta$ function form of strength A (coefficients of A and A multi-primes are used in this section throughout to denote undetermined constants). From Eqn. \ref{eq112}, it is easy to see that such a situation ensures a linear bath spectrum, well-known in the literature as the \textit{Ohmic Dissipation} \cite{weiss} paradigm.

\begin{eqnarray}
S(\omega) &=& M \omega \int _{0} ^{\infty} dt \delta (t) \cos \omega t \nonumber\\
S(\omega) &=& A M \omega
\label{eq116}
\end{eqnarray}

\noindent Thus, it can be concluded the usual Brownian Motion described by white noise and time-independent damping (a purely Markovian dynamics) is the result of a bilinear coupling to a bath which has a linear spectrum at the microscopic level imparting the fluctuations on the real physical trajectory of the particle.

\subsubsection{Non-Ohmic baths and noise correlations with memory}

\noindent The above theory has one drawback; the noise spectrum diverges at high frequencies. Now, for a realistic bath such a situation is untenable. The real problem lies in the fact that we need to consider a memory time for the noise correlations. Phenomenologically, this means that the two random kicks from the thermal bath have non-zero correlation at different times and the memory of the previous jolt persists for a certain amount of time (simply put, this is the inertial effect of the probe in a realistic medium). Of course, this time-scale should be small compared to the probe system dynamics and decay fast enough to ensure the impulsive nature of the stochastic thermal forces. Thus, the damping kernel is modified to include an exponentiated decay term,

\begin{eqnarray}
\gamma (t) &=& A \omega _{c}\exp (- \omega _{c} t)
\label{eq117}
\end{eqnarray}

\noindent effectively making the noise-correlations picking up a finite decay time-scale.

\begin{eqnarray}
\langle \Gamma(t)\Gamma (t')\rangle &=& A' \exp - \Big(\omega_{c} (t-t')\Big)
\label{eq118}
\end{eqnarray}

\noindent Noticeably, a very fast decay time-scale would correspond to a $\delta$-function and mimic a white noise. A straightforward integral on Eqn. \ref{eq112} immediately yields

\begin{eqnarray}
S(\omega) &=& \frac{A M \omega}{1+ \frac{\omega ^2}{\omega _{c} ^2}}
\label{eq119}
\end{eqnarray}

\noindent The equation above makes abundantly clear the role of Drude-like cut-off $\omega _{c}$ for the Lorenzian distribution. The unphysical divergence at high frequencies is also reguralised, leading to a realistic microscopic structure for the bath. Memory friction effects of the time-scale $\tau = \frac{1}{\omega _{c}}$ are introduced. The functional form of the Lorenzian also ensures that when relevant frequencies of the system are lower than the Drude cut-off $\omega_{c}$, the Ohmic approximation holds with a time-independent damping co-efficient dictating the open system dynamics.\\

\subsubsection{Super-Ohmic baths and temporally long-range noise}

\noindent Some words about super-Ohmic baths ($ S(\omega) \propto \omega ^ a$ and $a>1$) need to be addressed in some detail as this has significant bearing on the next section. Generally, such a spectral distribution is complemented by a suitable function that falls of with large $\omega$. Depending on systems and situations a sharp 
\begin{eqnarray}
S(\omega) &=& A \omega ^ a \Theta (1- \omega / \omega_ {c})
\label{eq120}
\end{eqnarray}

\noindent or a smooth exponentiated or Lorenzian (as in Eqn. \ref{eq119} for $a=1$) may be employed.

\begin{eqnarray}
S(\omega) &=& A \omega ^ a \exp \Big( - \omega / \omega _{c} \Big)
\label{eq121}
\end{eqnarray}

\noindent The cut-off frequency $\omega _c$ sets the fastest time scale of the dissipation dynamics (i.e. the fastest moving oscillator of the bath) and since this is a classical description of the problem, $\tau = \frac {1}{\omega _ {c}}$ only has a physical lower bound of $\tau >> \frac {\hbar}{K_B T}$.\\

\noindent In the course of our subsequent investigation, we explore the possibilities raised by regulated Super-Ohmic baths as espoused through the two equations \ref{eq120} and \ref{eq121}. Especially, we recall the real-time damping kernel and its relation to the noise correlations via the FDR is encapsulated in these forms of $S(\omega)$. Some interesting results are reported below. First, we take up the sharp-cutoff case as in Eqn \ref{eq120}. Using Eqn. \ref{eq114}, we find a compact form for the damping kernel,

\begin{eqnarray}
\gamma (t) = \Theta (t) \frac {2 \omega _c ^ a}{M \pi a} {}_{1}F_{2} \Big( \frac{a}{2}; \frac{1}{2},1+ \frac{a}{2}; -\frac{1}{4} \omega _c ^2 t^2 \Big)
\label{eq122}
\end{eqnarray}

\noindent with the introduction of the generalised Hypergeometric Function F. We take up some integer value cases exploring the nature of Super-Ohmic bath induced damping kernel.

\begin{align}
\gamma (t) = \Theta(t) \frac {2 \omega _c ^ 2}{ M \pi} t ^ {-2} \Big( \cos \omega _c t + t \sin \omega_c t -1 \Big) & ,a=2\\
\gamma (t) = \Theta(t) \frac {2 \omega _c ^ 3 }{ M \pi} t ^ {-3} \Big( 2 t \omega_c \cos \omega _c t - 2 \sin \omega_c t + \omega _c ^2 t^2 \sin \omega_c t \Big) & ,a=3
\label{eq123}
\end{align}\\

\noindent Next, we concentrate on the form of the spectral distribution with smooth exponentiated cut-off as in Eqn. \ref{eq121}. Through Eqn. \ref{eq114}, we find an exact form of the dissipation kernel for non-zero integer values of a and Super-Ohmic dictates that $a>1$:

\begin{eqnarray}
\gamma (t) &=& \Theta (t) \frac {2 \omega _c ^d}{M \pi} \Big(1+ t^2 \omega _c ^2 \Big) ^{-(a+1)} \cos \Big ((a+1) \tan ^{-1} \omega_c t \Big) \Gamma (a+1)
\label{eq124}
\end{eqnarray}

\noindent Here, $\Gamma (x)$ denotes the mathematical definition of a Gamma-function and the only condition for the existence of the integral is that $\frac{1}{\omega _{c}} >0$. Two representative values of a are studied explicitly to reveal the power-law structure of the kernel in $t$.

\begin{align}
\gamma (t) = \Theta (t) \frac {2 \omega _c ^2}{M \pi} \frac{1-\omega _c ^2 t^2}{(1+ \omega _c ^2 t^2)^2} &,a=2\\
\gamma (t) = \Theta (t) \frac {4 \omega _c ^3}{M \pi} \frac{1-\omega _c ^2 t^2}{(1+ \omega _c ^2 t^2)^3} &,a=3
\label{eq125}
\end{align}

\noindent Both the functions ( Eqns. \ref{eq122} and \ref{eq124}) exhibit power law decay in time and hence are examples of typical temporally long-range noise correlations. The difference between the two damping functions highlight the nature of cut-off introduced. Remarkably, cosine and sine terms dominate the short term correlations of the sharp cut-off case delaying the decay at large times.\\

\noindent Another instance of the Non-Ohmic bath can be extracted from noise-correlations which are long-range in temporal domain with a single exponent (the purist's version of nonequilibrium processes, like stochastic surface growth, etc). Such processes are well studied and over the years and covers a vast range of phenomena in nonequilibrium physics \cite{bar}. Here we use the the relation for noise correlations explicitly and use the inverted relation for $\gamma (t)$ to extract the functional form of the bath spectrum.

\begin{eqnarray}
\langle \Gamma(t)\Gamma (t')\rangle &=&  \frac{A}{|t-t'|^{\alpha -1}}
\label{eq126}
\end{eqnarray}

\noindent This produces a typically decaying power-law bath spectrum (using Eqn. \ref{eq114}) for values  $ 0<\alpha<1 $ (also, only in this range of $\alpha$ converges the relevant integral) that vanishes at large frequencies slowly with an exponent related to $\alpha $.

\begin{eqnarray}
S(\omega) &=& \frac {A M \Gamma (\alpha)}{\omega ^{\alpha - 1}} \cos(\frac{\alpha \pi}{2})
\label{eq127}
\end{eqnarray}

\noindent Though we adhere to a small range of $\alpha$, this doesnot complete the overall picture for long-range noises. It is obvious that for large time scales or small $\omega$, the damping kernel and the bath spectrum diverge. In such a case, a suitable low frquency cutoff has to be implemented setting the longest time-scale of the damping dynamics. Some famous growth models, like Edward-Wilkinson, Kardar-Parisi-Zhang, Eden, etc show such power law decay in the context of nonequilibrium growth processes of surfaces where standard deviation from the average height scales with exponents lying between 0 and 1 \cite{bar}. We should mention, however, in passing that a decaying power-law bath spectrum (with exponents greater than the above mentioned case) has serious physical problems like rendering $\gamma(0)$ as divergent, etc and hence Sub-Ohmic cases are not further pursued herein.\\

\noindent Therefore in this  careful study we realise that even under the gamut of quadratic bath, by using the powerful relations between the dissipation function and the bath noise spectrum, a variety of possibilities for the microscopic structure of the environment emerges and extensions to Gaussian non-white noise can be developed.

\subsection{Nonlinear baths and noise spectrum}

\noindent We concentrate on the quartic nonlinearity that has been introduced as a perturbative correction the bath mode potential. At zeroeth order, the standard FDR is retained making the noise correlations proportional to the temperature-independent standard damping function. Major corrections occur at the next perturbative order. This, together with the effect of weak coupling scheme between bath and system, is encapsulated in the FDR and the time nonlocal dissipation kernel \cite{bhadra1}.

\begin{eqnarray}
\langle \Gamma(t) \Gamma(t') \rangle &=& \frac{\gamma}{\beta} + \lambda^2 \epsilon
\sum_{\mu} \left[ \left\{ \frac{1}{2\beta} - \frac{1}{\beta^2} \frac{6b_{\mu}}{{(m_{\mu}{\omega_{\mu}}^2)}^2} \right\}  \right.
\nonumber \times \left. \frac{{C_{\mu}}^2}{m_{\mu} {\omega_{\mu}}^2} \cos \omega_{\mu} (t - t') \right].
\label{eq128}
\end{eqnarray}

\noindent As explored in our previous work \cite{bhadra1} in the weak system-bath coupling case, the above equation is moulded into form-invariant structure of the usual FDR through an effective damping kernel.

\begin{eqnarray}
\langle \Gamma(t)\Gamma (t')\rangle &=& \frac{{\gamma}^{[R]}(t-t')}{\beta}\nonumber
\label{eq129}
\end{eqnarray}

\noindent More precisely the renormalised form of the kernel can be written more precisely by implementing an effective form of the system-bath coupling that now scales with the energy scale set by the absolute temperature T.

\begin{eqnarray}
\gamma ^{[R]} (t-t^{'})&=& \lambda^{2}\sum_{\mu}\frac{C^{[R]2}_{\mu}}{m_{\mu} \omega_{\mu}^2} \cos[\omega_{\mu}(t-t')]
\label{eq130}
\end{eqnarray}

\noindent, The system-bath coupling parameter $C_{\mu}$ takes on an effective form:

\begin{eqnarray}
C^{[R]2}_{\mu} &=& C_{\mu} ^2 \left [ 1 + \epsilon \left(\frac{1}{2} - \frac{1}{\beta} \frac{6b_{\mu}}{{(m_{\mu}{\omega_{\mu}}^2)^2}} \right) \right]
\label{eq131}
\end{eqnarray}

\noindent However, the difference from the quadratic bath shows itself in the factor multiplying the perturbative parameter $\epsilon$. To adhere to the basic discretised definition of $S(\omega)$,

\begin{eqnarray}
S(\omega) &=& \frac{\pi}{2} \sum _{\mu} \frac{C_{\mu} ^2}{m_\mu \omega _{\mu}} \delta (\omega -\omega _\mu)
\label{eq132}
\end{eqnarray}
 
\noindent we observe that after taking the continuum limit for the bath modes, the general relation between the Fourier Transform of $\gamma (t-t')$ and $S(\omega)$ changes considerably. It reflects the perturbative correction to first order in the FDR and equivalently to the dissipation kernel while the condition $\epsilon =0$ limit gets back the previous form exactly.

\begin{eqnarray}
\tilde{\gamma}(\omega) &=& \lim _{\eta \to 0^+} \frac{-i \omega}{M} \frac{2}{\pi} \int _{0}^{\infty} d\omega ' \frac{S(\omega ')}{\omega '} \left[ 1 + \epsilon \left(\frac{1}{2} - \frac{1}{\beta} \frac{6b(\omega')}{m(\omega ')^2 \omega '^4} \right)\right]  \frac {1}{\omega '^2 - \omega ^2 + i\eta \omega} \nonumber\\
\label{eq133}
\end{eqnarray}

\noindent Nonlinearity of the bath induces terms in the formula that now depend on the internal structure of the bath explicitly: the mass and nonlinearity strength distributions, $m(\omega)$ and $b(\omega)$ (since they vary for each mode in the discretised case) and also on the temperature T. Since, the derivation of the real part of the damping kernel only involves calculating the complex integral around the pole around $\sqrt{\omega^2 - i\eta \omega}$, we  again get 

\begin{eqnarray}
\tilde{\gamma}' (\omega)  &=& \frac{S(\omega )}{M \omega } \left[ 1 + \epsilon \left(\frac{1}{2} - \frac{1}{\beta} \frac{6b(\omega)}{m(\omega )^2 \omega ^4} \right)\right] 
\label{eq134}
\end{eqnarray}

\noindent as the real part of Fourier Transformed damping kernel. The only condition that needs to be imposed for this derivation is that $b(\omega)$ and $m(\omega)$ donot have any non-trivial poles except at zero and $S(\omega)$ is not a pathological function, all very reasonable assumptions keeping in my mind that we are dealing with real physical systems. The inversion of this relation is also straightforward:

\begin{eqnarray}
S(\omega) &=& M \omega \left[ 1 + \epsilon \left(\frac{1}{2} - \frac{1}{\beta} \frac{6b(\omega)}{m(\omega )^2 \omega ^4} \right)\right] ^{-1} \int _{0} ^{\infty} dt \gamma(t) \cos \omega t
\label{eq135}
\end{eqnarray}\\

\subsubsection{Analysis of the emerging bath spectrum for a nonlinear bath}

\noindent The memoryless damping kernel, the hallmark of Markovian dynamics in open system, helps in evaluating the the spectral distribution of the bath directly.

\begin{eqnarray}
S(\omega) &=& A M \omega \left[ 1 + \epsilon \left(\frac{1}{2} - \frac{1}{\beta} \frac{6b(\omega)}{m(\omega )^2 \omega ^4} \right)\right] ^{-1} \int _{0} ^{\infty} dt \delta (t) \cos \omega t \nonumber\\
\label{eq136}
\end{eqnarray}

\noindent A range of possibilities for the spectrum can be studied depending on the functional form of the ration $\frac {b(\omega)}{m(\omega)^2}$. Let, the standard form of the ratio be taken as 

\begin{eqnarray}
\frac {b(\omega)}{m(\omega)^2} &=& A'' \omega ^ {(4 + \kappa)}
\label{eq137}
\end{eqnarray}

\noindent For the special case $\kappa =0$, 

\begin{eqnarray}
S(\omega) &=& A M \omega \left[ 1 + \epsilon \left(\frac{1}{2} - \frac{6 A''}{\beta} \right)\right] ^{-1}
\label{eq138}
\end{eqnarray}

\noindent Since, the form differs from the Ohmic and quadratic bath paradigm by a constant which is only dependent on temperature, we can envisage a situation where the mass $M$ of the probe particle is renormalised in line with similar discussions that arise in case of certain limtis of low-frequency approximations in Super-Ohmic baths \cite{weiss}. With a suitable cut-off defined at high frequencies (ensuring a memory kernel for the noise correlations), the exact formula for the bath spectrum is obtained.

\begin{eqnarray}
M_{eff} &=& M \left[ 1 + \epsilon \left(\frac{1}{2} - \frac{6 A''}{\beta} \right)\right] ^{-1}\\
S(\omega) &=& A M_{eff} \frac{\omega}{1+ (\frac{\omega _c}{\omega})^2}
\label{eq139}
\end{eqnarray} 

\noindent For $\kappa$ greater than zero, the emergence of a Super-Ohmic piece to the bath spectrum is observed for the Markovian limit where the distribution of bath modes assume the form

\begin{eqnarray}
S(\omega) &=& A M \omega \left[ 1 + \epsilon \left(\frac{1}{2} - \frac{1}{\beta} 6A'' \omega ^{\kappa} \right)\right] ^{-1}
\label{eq140}
\end{eqnarray}

\noindent adhereing to a temperature range that keeps the perturbative term in $\epsilon$ meaningful, we find

\begin{eqnarray}
S(\omega) &\approx & A M \omega \left[ \frac{1}{2} + \epsilon \left(\frac{1}{\beta} 6A'' \omega ^{\kappa} \right)\right]
\label{eq141}
\end{eqnarray}

\noindent Hence, under suitable Markovian approximation and range of parameters allowed by the perturbative parameter $\epsilon$ (ie having meaningful small values of the perturbative terms as a whole), we find a correction to the normal linear bath model and from the preceding discussions on Super-Ohmic bath one can draw similar conclusions about the nature of cut-off function regularising the spectral distribution of nonlinear baths. It is not pedantic to discuss the damping function here with a memory time scale $\omega_c$ involved,ie, showing non-Markovian characteristics. Using Eqn. \ref{eq135} and an exponentiated damping kernel, we find,

\begin{eqnarray}
S(\omega) &\approx & A M \frac {\omega}{1 + \frac{\omega ^2}{\omega _c ^2}} \left[ \frac{1}{2} + \epsilon \left(\frac{1}{\beta} 6A'' \omega ^{\kappa} \right)\right]
\label{eq142}
\end{eqnarray}

\noindent We find like unlike the quadratic case, fro $\kappa>1$, the spectral distribution is still unregularised at high frequencies. Almost in the same vain, a range of $\kappa$ , between zero and -1 produces a correction to the Ohmic model of Markovian dynamics resembing a soft Super-Ohmic bath, as distinctly revealed in the final form of the spectrum with an exponentiated cutoff function. Also, it helps to maintain the smallness of the perturbative corrections helping the expansion in $\epsilon$ meaningful.

\begin{eqnarray}
S(\omega) &=& A M \left[ \frac{1}{2} \omega + \epsilon \left(\frac{1}{\beta} 6A'' \omega ^{\kappa +1} \right)\right] \exp \Big( - \frac{\omega}{\zeta_c}\Big)
\label{eq143}
\end{eqnarray}

\noindent The cut-off function serves a dual role in the theory of nonlinear baths in a subtle manner. We astutely ignored the divergence arising from the distributions of $b(\omega )$ and $m(\omega )$ and their ratio which are automatically regularised at large frequencies due to the exponentiated decay of the natural frequencies of the mode oscillators themselves.

\subsection{Studying Velocity Correlations}

\noindent The last section paints a coarse-grained picture of a dissipative probe particle involving Langevin equation, memory-less damping kernel and Gaussian white noise. Sytem-bath Hamiltonian approach gives a Non-Markovian extension to the dynamics where the GLE emerges from a micrscopic picture. So, here we begin with a Generalised Langevin equation with an external potential, an effective damping kernel and nonlinear noise term, the later two serving as the basis for the robustness of a form-invariant FDR (Eqn. \ref{eq129}). Here we investigate the velocity correlations rigorously, arising in different cases of open system dynamics. 

\begin{eqnarray}
M\ddot{X} + V'(X(t)) + M \int_{0}^{t}\gamma ^{[R]} (t-t^{'})\dot{X}(t')dt^{'} &=&\Gamma(t)
\label{eq144}
\end{eqnarray}

\noindent We consider a harmonic potential with natural frequency $\Omega$. We remark that the counterterm ususally associated with a system-bath Hamiltonian serves to cancel out the renormalisation of the external potential in this formalism. However, in the quantum mechanical cases such terms do become important \cite{leg1}.

\begin{eqnarray}
V(X)&=&  \frac{1}{2} M \Omega ^2 X
\label{eq145}
\end{eqnarray}

\noindent The Generalised Langevin equation becomes,

\begin{eqnarray}
M\ddot{X} + M\Omega ^2 X + M \int_{0}^{t}\gamma ^{[R]} (t-t^{'})\dot{X}(t')dt^{'} &=&\Gamma(t)
\label{eq146}
\end{eqnarray}

\subsubsection {Over-damped Limit}

\noindent A suitable limit for evlauating diffusion processes is the over-damped condition  (due to Smoluchowski) where the inertial term is disregarded and the momentum is the fast variable of the dynamics. The dynamics is dominated by the damping kernel in this regime. From here, a lot of important characteristics of correlations of the probe particle can be derived and throws light on the nature of the effects of nonlinear baths.

\begin{eqnarray}
\Omega ^2 X + \int_{0}^{t}\gamma ^{[R]} (t-t^{'})\dot{X}(t')dt^{'} &=&\frac{\Gamma(t)}{M}
\label{eq147}
\end{eqnarray}

\noindent The damping kernel and its fourier transform (in discrete and continuum form) are laid out below with the usual definition of effective $C_\mu ^{[R]2}$.

\begin{eqnarray}
\gamma ^{[R]} (t-t^{'})&=& \lambda^{2}\sum_{\mu}\frac{C^{[R]2}_{\mu}}{m_{\mu} \omega_{\mu}^2} \cos[\omega_{\mu}(t-t')]\\
\tilde {\gamma} ^{[R]} (\omega) &=& \lim _{\eta \to 0^+} \frac{-i \omega}{M}\sum_{\mu}\frac{C^{[R] 2}_{\mu}}{m_{\mu} \omega_{\mu}^2} \frac {1}{\omega _{\mu} ^2 - \omega ^2 + i\eta \omega}
\label{eq148}
\end{eqnarray}

\noindent Utilizing the linear nature of the Langevin equation and the Fourier convolution theorem to untangle the damping integral, in terms of the dynamical variable $X$ and $\dot X$, the Fourier transform for Eqn. \ref{eq147} becomes,

\begin{eqnarray}
\Omega ^2 \tilde {X}(\omega) + \tilde{\gamma}^{[R]} (\omega) \tilde{V} (\omega)&=& \frac{\tilde{\Gamma}(\omega)}{M}
\label{eq150}
\end{eqnarray}

\indent The following identities and formulae help us to evaluate dynamical correlations in term of the Fourier transformed coordinates in the above equation.

\begin{eqnarray}
\tilde {X}(\omega)&=& \int dt \exp (i\omega t) X(t)\nonumber\\
V(t)&=& \dot {X}(t)\nonumber\\
\tilde {V}(\omega)&=& -i\omega \tilde {X}(\omega)
\label{eq151}
\end{eqnarray}

\noindent Thus concentrating on the velocity of the overdamped probe particle in a nonlinear bath, we find the following equations,

\begin{eqnarray}
\tilde {V}(\omega)&=& \frac{1}{M} \frac{\Gamma(\omega)}{\tilde{\gamma}^{[R]} (\omega) + i\frac {\Omega^2}{\omega}}\\
\tilde {V}(\omega)\tilde {V}(\omega')&=& \frac{1}{M^2} \frac{\Gamma(\omega)\Gamma(\omega')}{(\tilde{\gamma}^{[R]} (\omega) +i \frac {\Omega ^2}{\omega })(\tilde{\gamma}^{[R]} (\omega') +i \frac {\Omega ^2}{\omega '})}
\label{eq152}
\end{eqnarray}

\noindent Next we evaluate the correlations $\langle..\rangle$ in a straightforward manner  and find that dynamical correlations are directly proportional to the noise-correlations when F.T. coordinates are used. 

\begin{eqnarray}
\langle \tilde {V}(\omega)\tilde {V}(\omega') \rangle &=& \frac{1}{M^2} \frac{\langle \Gamma(\omega)\Gamma(\omega')\rangle}{(\tilde{\gamma}^{[R]} (\omega) +i \frac {\Omega ^2}{\omega })(\tilde{\gamma}^{[R]} (\omega') +i \frac {\Omega ^2}{\omega '})}
\label{eq153}
\end{eqnarray}

\noindent Now, the noise-correlations have been evaluated in detail previously for a nonlinear bath giving rise to an effective damping kernel. In a compact form (with the mass term incorporated), they become (see Appendix for details)

\begin{eqnarray}
\langle \Gamma(t)\Gamma (t')\rangle &=& \frac{{M \gamma}^{[R]}(t-t')}{\beta}\\
\langle \Gamma(\omega)\Gamma(\omega')\rangle &=& \frac{M Re \left[\tilde{\gamma}^{[R]} (\omega)\right] \delta (\omega + \omega')}{\pi \beta}
\label{eq154}
\end{eqnarray}

\noindent Thus, the velocity correlations become,
\begin{eqnarray}
\langle \tilde {V}(\omega)\tilde {V}(\omega') \rangle &=& \frac{1}{M \pi \beta} \frac{Re \left[\tilde{\gamma}^{[R]}(\omega)\right] \delta (\omega + \omega')}{(\tilde{\gamma}^{[R]} (\omega) +i \frac {\Omega ^2}{\omega })(\tilde{\gamma}^{[R]} (\omega') +i \frac {\Omega ^2}{\omega '})}
\label{eq155}
\end{eqnarray}

\noindent Finally, the double inverse Fourier Transform would lead to the time correlation for the velocity of the probe particle. Notably, one of the integrals is killed off by the presence of a $\delta$-function.

\begin{eqnarray}
\langle V(t) V(t') \rangle &=&  \frac{1}{M \pi \beta} \int \int d\omega d\omega ' \exp (-i\omega t)\exp (-i\omega ' t')\langle \tilde {V}(\omega)\tilde {V}(\omega') \rangle \nonumber\\
&&=\frac{1}{M \pi \beta} \int d\omega \exp (-i\omega (t-t')) \frac{Re \left[\tilde{\gamma}^{[R]} (\omega)\right]}{(\tilde{\gamma}^{[R]} (\omega) +i \frac { \Omega ^2}{\omega })(\tilde{\gamma}^{[R]} (-\omega) -i \frac {\Omega ^2}{\omega})}\nonumber\\
\label{eq156}
\end{eqnarray}\\

\noindent This is the master formula via which we should be able to ideally determine dynamical correlations in the over-damped limit for generic noise correlations, generalised nonlinear baths and evaluate different limits for the parameters present in such open dynamics problem. However, as we will see the integrals throw significant challenges and certain methods are deemed unsuitable even when they make the calculations easier. We would like to point out at this very juncture that such Fourier Transform methods are inspired by the works of Dhar et.al. on Anomalous Heat Conduction and Diffusion in low-dimensional system \cite{dhar4}. There, though the problem is essentially nonequilibrium and the main focus remains the evaluation of a nonequilibrium steady state, such system-bath dynamics formalism provides the shortest route to calculating correlations of different kinds with various physical implications. Yet, we note, that those problems still adhere to the scenario of quadratic baths and hence long-time correlations in the bulk mimiccharacteristics of thermal nature.

\subsubsection{Quadratic baths and Markovian dynamics}
\noindent To get a handle on the complicated integral in the R.H.S. of Eqn. \ref{eq156}, we first try out the simplest case, that of a free probe particle in a qudratic thermal environment. This means, in the master formula the following conditions are true.

\begin{eqnarray}
\epsilon &=0,  & \Omega =0\nonumber\\
\end{eqnarray}

\noindent The master formula Eqn. \ref{eq156} boils down to

\begin{eqnarray}
\langle {V}(t) V(t') \rangle &=& \frac{1}{M \beta} \int d\omega \exp (-i\omega (t-t'))(\tilde{\gamma})^{-1}(-\omega)
\label{eq157}
\end{eqnarray}

\noindent with an Ohmic approximation leading to a frequency independent dissipative kernel,

\begin{eqnarray}
\tilde {\gamma}(\omega)&=& \gamma
\label{eq158}
\end{eqnarray}

\noindent The velocity correlations for the Markovian dynamics becomes

\begin{eqnarray}
\langle {V}(t)V(t') \rangle &=& \frac{1}{M \gamma \beta} \delta (t-t')
\label{eq159}
\end{eqnarray}

\noindent Such a result can easily be checked by going back to section 1 of this paper and solving for velocity in Eqn. \ref{eq100} without the inertial term (the over-damped) limit. The co-efficient of the delta-function noise for a general case is of course chosen arbitrarily and the equipartition theorem asserts its value in this case. Physically, in the heavy damping regime the velocity of a free probe particle is completely dictated by the thermal noise forces itself and is defined as a pure Markov process.\\

\noindent Next we consider the case of a Simple Harmonic Oscillator as the probe system.

\begin{eqnarray}
\Omega &\neq& 0 \nonumber
\end{eqnarray}

\noindent In the limit of Ohmic approximation and frequency-independent dissipation function, we find

\begin{eqnarray}
\langle {V}(t) V(t') \rangle &=& \frac{1}{\beta} \int d\omega \exp (-i\omega (t-t')) \frac{\gamma}{\gamma^2 \omega ^2 + \Omega ^4}\nonumber\\
\langle {V}(t) V(t') \rangle &=& \frac{1}{2 M \beta \Omega ^2 } \exp \Big(- \frac{\Omega ^2}{\gamma} |t-t'| \Big) 
\label{eq160}
\end{eqnarray}

\noindent The above formula is valid for all $t \neq t'$. The correlation is well regularised at very short times and large time decay point to the relaxation mechanism induced by the minima of the harmonic potential on the damped particle.

\subsubsection {Free particle with inertial term}

\noindent A small discussion about the full GLE in Eqn. \ref{eq146} is warranted at this point for $\Omega =0$ case due to its similar structure to the preceding discussion. The general velocity correlation under such a paradigm becomes,

\begin{eqnarray}
\langle {V}(t) V(t') \rangle &=& \frac{1}{M \beta} \int d\omega \exp (-i\omega (t-t')) \frac{\tilde{\gamma} (\omega)}{(\tilde{\gamma}(\omega) +i \omega)(\tilde{\gamma} (-\omega) -i \omega)}
\label{eq161}
\end{eqnarray}

\noindent With the Ohmic approximation in place, the dynamical velocity correlation mimics the form in Eqn. \ref{eq160},

\begin{eqnarray}
\langle {V}(t) V(t') \rangle &=& \frac{1}{2 M \beta}\exp \Big(- \gamma |t-t'| \Big)
\label{eq162}
\end{eqnarray}

\subsubsection {For Coloured Spectrum of Non-linear baths}

\noindent The relevant quantity is the Fourier Transformed form of the effective dissipation function

\begin{eqnarray}
\tilde{\gamma}^{[R]}(\omega)=-\frac{i\omega}{M}\sum_\mu \frac{C^{[R]2}_\mu}{m_\mu \omega_\mu^2}\frac{1}{\omega_\mu^2-\omega^2}
\label{eq162}
\end{eqnarray}

\noindent In the continuum limit this becomes
\begin{eqnarray}
\tilde{\gamma}^{[R]}(\omega)=-\frac{i\omega}{M} \int_{0}^{\infty}d\omega'\frac{C^{[R]2}(\omega')}{m\omega'^2}\frac{1}{\omega'-\omega^2}
\label{eq163}
\end{eqnarray}

\noindent Now,
\begin{eqnarray}
C^{[R]2}(\omega')=C^2[1+\epsilon(\frac{1}{2}-\frac{1}{\beta}6\omega'^ \kappa)]\\
where &&\omega'^ \kappa=\frac{b(\omega)}{m(\omega)^2}\nonumber
\end{eqnarray}

\noindent and therefore,

\begin{eqnarray}
\tilde{\gamma}^{[R]}(\omega)=-\frac{i\omega}{M}\int d\omega'(\frac{\alpha}{\omega'^2}+\epsilon\alpha'\omega'^{\kappa-2})\frac{1}{\omega'^2-\omega^2}
\label{eq164}
\end{eqnarray}

\noindent Also we have shown previously that the spectral function S($\omega$) can be written as
\begin{eqnarray}
S(\omega)=AM\omega(\frac{1}{2}+\epsilon6 A''\omega^ \kappa)
\label{eq165}
\end{eqnarray}

\noindent So if we can write down the spectral function due to non-linear bath as polynomial in $\omega$ then we can find the respective expressions for $\tilde{\gamma}^{[R]}(\omega)$ and S($\omega$) and put them in the master formula for calculation of velocity correlation( Eqn. \ref{eq156}) and obtain an exact form for the velocity correlations. The choice $\kappa =0$ gives back the Markovian results for the quadratic case as in Eqn. \ref{eq159}, \ref{eq160} and \ref{eq161} with the added ingredient that all the probe mass terms become $M_{eff}$. That is, an effective mass of the probe particle is introduced due to the nonlinearity as discussed in the previous sections.

\newpage \section{Nonequilibrium Evolution and Correlations}

\subsection {Some opening remarks on Non-equilibrium dynamics}
The system-reservoir approach has been used extensively in various fields of Physics and Chemistry and the main premise of the bipartite nature of a system interacting with its surrounding has been the basis for many others. Nevertheless, one of its  profound applications still remains in the formulation of classical Brownian Motion in an equilibrium setting as well as its quantum counterpart. In this context, non-Markovian dynamics, decoherence effects, Kramer's rate over an energy barrier, etc have been studied extensively via the introduction of non-linear external potentials and non-linear system-reservoir couplings. Non-equilibrium dynamics,however,still remains devoid of any such all-encompassing formalism. But, remarkable progress have been made over the years, especially due to Linear Response theory, to understand the phenomenological implications of systems away from equilibrium. Many other rich avenues stand on their own due to far-reaching insights in Hydrodynamics, special Integrable Models, to name a few. But a system-reservoir Hamiltonian approach leading to Langevin Equations and other familiar counterparts of equilibrium calculations, though, remain few and far in recent literature. Nevertheless, an ongoing study where such  an approach is abundant is in the field of Heat Conductivity \cite{dhar4}.The phenomenon is inherently non-equilibrium and the standard system-reservoir formalism (the heat conductor being the system and connected to two Langevin baths at different temperature) has been shown to yield remarkably accurate results concerning the anomalous thermal conductivity in low-dimensions. Recently, a Linear Response approach also casts the problem in a new perspective \cite{liu}; it shows the effect of a small energy perturbation on a Hamiltonian system and tracks the evolution of this excess energy through the thermal conductor in probabilistic terms leading to precise predictions.\\

\noindent Since, in previous works \cite{bhadra1} nonlinear baths have been shown to naturally give rise to non-equilibrium correlations (both in terms of bath correlations and dynamical averages), our main aim of this section is to find such \textit{non-equilibrium effects due to small non-linear perturbations (Linear Response regime) on a simple quadratic thermal reservoir}. The framework is rooted in the system-reservoir description and we envisage a situation where the perturbation acts on the bath before t=0 at which point it is suddenly switched off. The resulting evolution of the system leaves its signature in the non-equlibrium correlations of noise, velocity and other relevant physical observables. The interplay of non-linearity and perturbation on the bath-modes is our primary concern here. We give explicit formulae for the first and second moments of the thermal noise which are now averages taken over a non-equilibrium ensemble. The noise-noise correlation, that encodes these non-equilibrium corrections to the standard FDR, follows naturally. We also investigate a specific case: a quartic perturbation. The fluctuations persist for long times with the essential safety of being bounded and periodic since after all we use a perturbative treatment and the bath modes are pinned. We emphasise, that throughout the study, we work in the Linear Response regime and employ perturbative techniques; large perturbations are not ammenable to such a treatment.

\subsection{Recap of the system-bath dynamics with quadratic and nonlinear modes}

\noindent A system-reservoir Hamiltonian in an equilibrium setting involves four separate terms: the system, the reservoir, the coupling between the two and a counter-term.
\begin{eqnarray}
H_{total}&=&H_{S}+H_{R}+H_{SR}+H_{CT}\nonumber\\
H_{S}&=&\frac{P^{2}}{2M}+V(X)\nonumber\\
H_{R}&=&\sum_{\mu=1}^{N} \frac{p_{\mu}^{2}}{2m_{\mu}}+\sum_{\mu=1}^{N}\frac{m_{\mu}^{2}\omega_{\mu}^{2} q_{\mu}^{2}}{2}\nonumber\\
H_{SR}&=& -\sum_{\mu=1}^{N} \lambda C_{\mu}q_{\mu}X
\label{eq300}
\end{eqnarray}

\noindent The Lagrange's equation of motion follows naturally and with a little algebra we can establish a Generalised Langevin Equation(GLE) with the damping co-efficient and noise term suitably identified.
\begin{eqnarray}
M\ddot{X}&=& -V^{'}(X) -\int_{0}^{t}\gamma(t-t^{'})\dot{X}dt^{'} +\Gamma(t)\nonumber\\
\gamma(t-t^{'})&=&\lambda^{2}\sum_{\mu}^{N}\frac{C_{\mu}^{2}}{m_{\mu}^{2}\omega_{\mu}^{2}}\cos{\omega_{\mu}(t-t^{'})}\nonumber\\
\Gamma(t)&=&\sum_{\mu}^{N}C_{\mu}[q_{\mu}(0)\cos{\omega_{\mu}(t)} + \frac{p_{\mu}(0)}{m_{\mu}\omega_{\mu}}\sin{\omega_{\mu}(t)}]
\label{eq301}
\end{eqnarray}

\noindent With a initial thermal distribution for the set of bath oscillators , $P_{eq}=\frac{1}{Z_{eq}} exp\Big(-\beta H_{R}\Big)$, and an initial slippage term \cite{weiss}, the FDR of the second kind is established:
\begin{eqnarray}
\langle \Gamma (t) \rangle &=& 0 \nonumber\\
\langle \Gamma (t)\Gamma (t^{'})\rangle &=& \frac{\gamma (t-t^{'})}{\beta}
\label{eq302}
\end{eqnarray}

\noindent On an addition of a quartic nonlinearity parametrised by the perturbation parameter $\lambda$, the bath Hamiltonian becomes

\begin{eqnarray}
H_R' = \sum_{\mu} \left[ \left\{\frac{{p_{\mu}}^2}{2m_{\mu}} + \frac{1}{2}m_{\mu} {\omega_{\mu}}^2 {q_{\mu}}^2 \right\}
+ \epsilon b_{\mu} {q_{\mu}}^4 \right]
\label{eq303}
\end{eqnarray}

\noindent With the definitions of Eqs.(\ref{eq300}) and (\ref{eq303}), the canonical distribution function for the initial spectrum of the bath variables takes the form

\begin{eqnarray}
P (\{q_{\mu}(0)\}, \{p_{\mu}(0)\}) &=& Z^{-1} \exp -\beta \big[\sum_{\mu} \big(\frac{{p_{\mu}}^2(0)}{2m_{\mu}} +  \frac{1}{2}m_{\mu} {\omega_{\mu}}^2 {q_{\mu}}^2(0)\big) \nonumber\\
&+& \epsilon b_{\mu} {q_{\mu}}^4(0) \big]\nonumber\\
\label{eq304}
\end{eqnarray}

\noindent Our detailed treatment in reference \cite{bhadra1} helps us in writing down the noise-correlations (i.e. the extended form of the FDR) in a rigorous fashion. 

\begin{equation}
\langle \Gamma(t) \rangle = 0.
\label{eq305}
\end{equation}

\noindent The vanishing of the first moment of the noise, in keeping with our usual experience, happens because of the symmetry in the reservoir potential \cite{bhadra1}.\\ 

\noindent Defining a suitable quantity,

\begin{equation}
D_{\mu} = \frac{1}{2} - \frac{1}{\beta} \frac{6b_{\mu}}{{(m_{\mu}{\omega_{\mu}}^2)}^2}.
\label{eq306}
\end{equation},

\noindent the FDR (of the second kind) for the quartic bath takes the explicit form,

\begin{eqnarray}
\langle \Gamma(t) \Gamma(t') \rangle &=& \frac{\gamma}{\beta} + \lambda^2 \epsilon
\sum_{\mu} \left[ \left\{ \frac{1}{2\beta} - \frac{1}{\beta^2} \frac{6b_{\mu}}{{(m_{\mu}{\omega_{\mu}}^2)}^2} \right\}  \right.
\nonumber \\
&\times& \left. \frac{{C_{\mu}}^2}{m_{\mu} {\omega_{\mu}}^2} \cos \omega_{\mu} (t - t') \right].
\label{eq307}
\end{eqnarray}

\noindent In parallel to our previous work in the weak system-bath coupling case, the above equation is moulded into form-invariant structure of the usual FDR through an effective damping kernel.

\begin{eqnarray}
\langle \Gamma(t)\Gamma (t')\rangle &=& \frac{{\gamma}^{[R]}(t-t')}{\beta}\nonumber\\
\label{eq308}
\end{eqnarray}

\subsection{A quenched initial distribution and Louiville operator}
With the specifics in place, we now concentrate on a formalistic derivation of a non-equilibrium scenario. We consider a small spatial perturbation $H_{1R}$ on the reservoir Hamiltonian that has been acting till t=0 and can be identified with Equation (\ref{eq303}).  
\begin{eqnarray}  
H_R^{'}&=&H_{R}+\epsilon H_{1R}
\label{eq309}
\end{eqnarray}

\noindent Then the perturbation is suddenly switched off with the reservoir Hamiltonian reduced to the simple quadratic Hamiltonian $H_{R}$. We now want to calculate the effect of such a small perturbation on the noise-correlations for all subsequent times. Limiting ouselves to the Linear Response regime, we evaluate the non-equilibrium partition function $Z$ for the reservoir co-ordinates perturbatively and the associated probability distribution in the small $\epsilon$ limit.
\begin{align}
Z^{'}&= \int exp \Big(-\beta H_R^{'}\Big)d\Omega\nonumber\\
&=\int exp \Big(-\beta(H_{R}+\epsilon H_{1R})\Big)d\Omega\nonumber\\
&\approx\int \Big(1-\beta \epsilon  H_{1R}\Big)exp \Big(-\beta H_{R}\Big)d\Omega\nonumber\\
&=Z\Big(1-\beta \epsilon \langle H_{1R} \rangle_{eq}\Big)
\label{eq310}
\end{align}\\
where,
\begin{align}
d\Omega&\equiv \Big(\prod_{\mu =1}^{N}\prod_{\mu =1}^{N}dp_{\mu}dq_{\mu}\Big)
\label{eq311}
\end{align}

\noindent Physically this means, an out-of-equilibrium state is prepared at $t=0$ by switching off the quenched Hamiltonian $H_R{'}$,which is assumed to have acted from the infinite past. In different light, this is an initial-value problem with an initial probability distribution in a displaced,frozen-equilibrium ensemble probability, whose future time evolution depends on an unperturbed Louivillean $L$ \cite{liu}. The expansion for a near-equilibrium scenario perturbatively leaves only the effect of \textit{equilibrium correlations} as expected from such a theory and becomes the bedrock of the following analysis. However, several steps follow before a truly first order expression for the correlations can be found.\\

\noindent The nonequilibrium evolution of the probablity distribution is governed by the Louivillean,
\begin{eqnarray}
P_{R}^{neq}(t)&= e^{tL_{total}}P_{R}^{neq}(0)
\label{eq312}
\end{eqnarray}

\noindent where the total Louivillean contains the coupled system-bath Hamiltonian as a whole.

\begin{eqnarray}
L_{total}&=&  \left\{H_{S}+H_{R}+H_{SR}+H_{CT}, ...\right\}
\label{eq313}
\end{eqnarray}

\noindent $\left\{..,..\right\}$ is the usual Poisson Bracket with respect to initial bath co-ordinates. Now, the system and counter-term Hamiltonian are only dependent on the probe co-ordinates and the Poisson Bracket with respect to the initial quenched bath probability distribution is zero. Also, we recall that $H_{SR}$ is controlled by a perturbative parameter $\lambda$ (weak system-bath coupling approximation) and hence an expansion leads to,

\begin{eqnarray}
P_{R}^{neq}(t)&= e^{tL_{R}}\Big[P_{R}^{neq}(0) + \lambda t \left\{H_{SR},P_{R}^{neq}(0)\right\} + O(\lambda ^2 t^2)\Big]
\label{eq314}
\end{eqnarray}

\noindent The above analysis puts a bound on the time interval $(t<< \frac{1}{\lambda})$ on which subsequent evaluation of correlations remain meaningful under a perturbative approximation. Physically, the above limit indicates the presence of nonequilibrium correlations on time scales much less than that of dissipation of the probe system dynamics since the damping kernel is controlled by the parameter $\lambda$ primarily [Eq. \ref{eq301}]. A detailed analysis of the two terms follow.\\

\subsection {Evaluation of the first term}

\noindent The equation above has been written down till first order in $\lambda$. We evaluate term by term the explicit formulae for the expressions on the R.H.S. upto that order. Using the result for an expanded partition function of the quenched distribution at $t=0$, we arrive at

\begin{align}
P_{R}^{neq}(t)&= e^{tL_{R}}P_{R}^{neq}(0)\nonumber\\
&=\frac{1}{Z^{'}}e^{tL_{R}}e^{-\beta \epsilon H_{1R}}e^{-\beta  H_{R}}\nonumber\\
&\approx\frac{1}{Z}\Big(1+\epsilon \beta \langle H_{1R} \rangle _{eq}\Big)e^{tL_{R}}\Big(1-\beta \epsilon H_{1R}\Big)e^{-\beta  H_{R}}\nonumber\\
&=e^{tL_{R}}(1-\beta \epsilon \Delta H_{1R})P_{R}^{eq}+O(\beta^{2})
\label{eq315}
\end{align}

where,
\begin{eqnarray}
\Delta A&\equiv A- \langle A \rangle _{eq}
\label{eq316}
\end{eqnarray}

\noindent a standard quantity that measures statistical fluctuations of a physical variable when perturbed slightly out of equilibrium.\\

\noindent The next natural step is to find the non-equilibrium averages of the thermal noise for times $t>0$.The one-point average, which is simply zero for unbiased noise, will pick up corrections due to the perturbation of the bath Hamiltonian. We remark, that the noise term we consider is of the unperturbed Hamiltonian. For a memory-less damping kernel, such an assumption is well justified as the noise remembers no previous memory of the perturbation for  $t>0$. Using eqns \ref{eq310} and \ref{eq315},
\begin{align}
\langle \Gamma(t) \rangle _{neq}&=\int\Gamma(t)P_{R}^{neq}(\Omega,t)d\Omega\\
&=\langle \Gamma(t) \rangle _{eq} - \beta \epsilon \int\Gamma(t)e^{tL_{B}}P_{R}^{eq}(\Omega)\Delta H_{1R}d\Omega\nonumber\\
&=\langle \Gamma(t) \rangle _{eq}-\beta \epsilon \langle \Gamma(t)\Delta H_{1R}(0) \rangle _{eq}
\label{eq317}
\end{align}

\noindent The Fluctuation-Dissipation Relation is also affected by the perturbation as now the averages are calculated over the non-equilibrium probability distribution. For this we need a product-separable Probability function for the primed and unprimed co-ordinates without any loss of generality. 
\begin{align}
P_{R}^{neq}(\Omega,\Omega{'},t,t{'})&=P_{R}^{neq}(\Omega,t)P_{R}^{neq}(\Omega{'},t^{'})
\label{eq318}
\end{align}

\noindent A similar expansion via eqn. \ref{eq310} and \ref{eq315} gives
\begin{align}
P_{R}^{neq}(\Omega,\Omega{'},t,t{'})&\approx\Big(P_{R}^{eq}(\Omega)- \beta \epsilon e^{tL_{B}}\Delta H_{1R}(\Omega)P_{R}^{eq}(\Omega)\Big)\nonumber\\
&\times \Big(P_{R}^{eq}(\Omega^{'})- \beta \epsilon e^{t^{'}L_{B}}\Delta H_{1R}(\Omega^{'})P_{R}^{eq}(\Omega^{'})\Big)\nonumber\\
 &=P_{R}^{eq}(\Omega)P_{R}^{eq}(\Omega{'}) - \beta \epsilon \Big(e^{tL_{B}}P_{R}^{eq}(\Omega)\Delta H_{1R}(\Omega)P_{R}^{eq}(\Omega{'}\Big)\nonumber\\
& -\beta \epsilon \Big(e^{t^{'}L_{B}}P_{R}^{eq}(\Omega^{'})\Delta H_{1R}(\Omega{'})P_{R}^{eq}(\Omega)\Big) + O(\beta^{2}) \nonumber\\
\label{eq319}
\end{align}

\noindent The final evaluation of the 2-point noise correlation leads to

\begin{align}
\langle \Gamma(t)\Gamma(t^{'}) \rangle _{neq}&=\int\!\!\!\int d\Omega d\Omega^{'} P_{R}^{neq}(\Omega,\Omega{'},t,t{'}) \Gamma(t) \Gamma(t^{'})\\
&\approx\int\!\!\!\int d\Omega d\Omega^{'}\Gamma(t)\Gamma(t^{'})\Big(P_{R}^{eq}(\Omega)P_{R}^{eq}(\Omega{'})\nonumber\\
& - \beta \epsilon \Big(e^{tL_{B}}P_{R}^{eq}(\Omega)\Delta H_{1R}(\Omega)P_{R}^{eq}(\Omega{'})+e^{t^{'}L_{B}}P_{R}^{eq}(\Omega^{'})\Delta H_{1R}(\Omega{'})P_{R}^{eq}(\Omega)\Big)\nonumber\\
& + O(\beta^{2})\Big)\nonumber\\
 &= \langle \Gamma(t)\Gamma(t^{'}) \rangle _{eq}-\beta \epsilon \langle \Gamma(t)\Gamma(t^{'})\Delta H_{1R}(0)) \rangle _{eq}\nonumber\\
& + (t,t{'})+O(\beta^{2})
\label{eq320}
\end{align}

\noindent We now consider a specific non-linear perturbation to the thermal bath before linear dynamics commences. It is quartic in the reservoir co-ordinates and the simplest stable addition to the Hamiltonian.

\begin{align}
H_R^{'}&=H_{R}+H_{1R}\nonumber\\
&=\sum_{\mu=1}^{N} \frac{p_{\mu}^{2}}{2m_{\mu}}+\sum_{\mu=1}^{N}\frac{m_{\mu}^{2}\omega_{\mu}^{2} q_{\mu}^{2}}{2}+ \epsilon\sum_{\mu=1}^{N}a_{\mu} q_{\mu}^{4}
\label{eq321}
\end{align}

\noindent The parameter $\epsilon$ is the small co-efficient around which any viable perturbative expansion can be done. For the sake of generality, all bath modes are perturbed initially as is clear from the fact that the summation runs over all N-oscillators in the second term of the R.H.S. of the above equation. A more realistic situation would be to consider a random distribution of modes being disturbed; this however will not affect the general results we lay down below. Also, as we have elucidated earlier, this perturbation acts till $t=0$ after which it switches off and the system-reservoir Hamiltonian evolves under the influence of the unperturbed bath $H_{R}$. The first moment of the noise calculated for a quartic perturbation remains zero in the Linear Response regime. This result essentially relies on the fact that the System-Bath Hamiltonian is an even function in the bath co-ordinates (and the noise is an odd function of the same.) Let the noise term for the dynamics under a quadratic bath Hamiltonian be denoted by $\Pi (t)$ from hereon.

\begin{eqnarray}
\Pi (t) &=& \sum_{\mu}^{N}C_{\mu}[q_{\mu}(0)\cos{\omega_{\mu}(t)} + \frac{p_{\mu}(0)}{m_{\mu}\omega_{\mu}}(t=0)\sin{\omega_{\mu}(t)}]
\label{eq322}
\end{eqnarray}

\noindent The averages on the other hand are calculated with the corresponding expansion for the nonequilibrium time-evolving probability distribution in Eqn. \ref{eq315}. Its easy to show,
\begin{align}
\langle \Pi_{1} (t) \rangle ^1 _{neq}&=0
\label{eq323}
\end{align}

\noindent with the superscript denoting the coarse-graining with respect to the first term in Eqn (\ref{eq314}).\\

\noindent Furthermore, the noise-noise correlation, as calculated in Eqn. (\ref{eq320}), leads to the non-equilibrium counterpart of the standard FDR. The FDR is the mathematical equivalence of the two main aspects of Brownian motion: the random noise and molecular dissipation. Physically, a perturbation in the thermal bath modes should translate to corrections to the damping kernel also. We find exactly such a scenario in our formulation. A factor of 2 has been multiplied to accommodate the $(t,t{'})$ symmetric term.

\begin{align}
\langle \Pi_{1} (t)\Pi_{1} (t{'}) \rangle ^1 _{neq} &= \sum_{\mu}^{N}\frac{\lambda^{2}}{\beta}\left[1+\epsilon \left( \frac{6a_{\mu}}{\beta m_{\mu}^{2}\omega_{\mu}^{4}}-1 \right)\right]\frac{C_{\mu}^{2}}{m_{\mu}\omega_{\mu}^{2}}\cos{\omega_{\mu}(t-t^{'})}\nonumber\\
\label{eq324}
\end{align}

\noindent Eqns (\ref{eq323}) and (\ref{eq324}) complete the evaluation via the first term in Eqn. (\ref{eq314}). These equations are general enough to accommodate for any small and stable spatial perturbation to the thermal bath. As a hallmark of the  linear response regime,the equations show that the non-equilibrium corrections to the equilibrium average of the noise-term involves only the equilibrium correlations between the noise and the perturbation Hamiltonian. The structure of these correlations encode the information about how the small disturbance to the thermal bath affects the noise correlations in the subsequent times. Again, due to the explicit involvement of the perturbation term $(H_{1R})$ in these correlations, its non-linear form also plays a crucial role in determining the non-equilibrium effects, as we will see in the next section. \\

\subsection {The second term: evaluating the Poisson bracket}

\noindent The Poisson bracket in the second term of the expanded Louivilliean, Eqn. (\ref{eq314}), explicitly can be written as

\begin{eqnarray}
\left\{H_{SR}, P_{neq}(0) \right\}_{p_{\mu}(0),q_{\mu}(0)} &=& -\lambda \Big [ \frac{\delta H_{SR}}{\delta q_{\mu}(0)} \frac{\delta P_{neq}(0)}{\delta p_{\mu}(0)} - \frac{\delta H_{SR}}{\delta p_{\mu}(0)} \frac{\delta P_{neq}(0)}{\delta q_{\mu}(0)} \Big]\nonumber\\
\label{eq325}
\end{eqnarray}

\noindent Recalling the expansions of the partition function and quenched distribution of initial bath modes from Eqns. (\ref{eq315}) and (\ref{eq325}), and using the simple linear solution for the bath modes for $t>0$,

\begin{eqnarray}
q_{\mu}(t) &=& q_{\mu} (0) \cos (\omega _{\mu} t) + \frac{p_{\mu} (0)}{m_{\mu} \omega _{\mu}} \sin (\omega _{\mu}t) \nonumber\\
&&+ \frac{\lambda}{m_{\mu} \omega _{\mu}} \int_{0}^{t} dt{'}\sin (\omega _{\mu}(t-t{'}))X(t{'})
\label{eq326}
\end{eqnarray}

\noindent we arrive at a rigorous formula for the Poisson Bracket:

\begin{eqnarray}
\left\{H_{SR}, P_{neq}(0) \right\} &=& - \lambda X(t) \frac{1}{Z} (1+ \epsilon \beta \langle H_{1R} \rangle ) \Big[ \sum _{\mu} \cos (\omega _{\mu} t) \frac{p_{\mu} (0)}{m_{\mu} \omega _{\mu}}  \exp (-\beta H_{R})\nonumber\\
&+& \sum _{\mu} \sin (\omega _{\mu} t) m_{\mu} \omega _{\mu} ^2 q_{\mu}(0) \exp (-\beta H_{R}) \Big] 
\label{eq327}
\end{eqnarray}

\noindent The full machinery of the expansion due to the small nonlinearity parameter $\epsilon$ is exploited in the above calculation leaving only a quadratic Hamiltonian based probability distribution to deal with. Several remarks are in order. The presence of the system-bath coupling (weak limit) explicitly in the evolution dynamics of the probability kernel signals the dependence on the dynamics of the probe system; such phenomena are common in nonlinear Langevin dynamics showing nonequlibrium features \cite{maes1,maes3,maes4}. In a way, the back-reaction of the dynamics of the probe is felt for short times by the bath if a quenched initial condition is met. Since the above scenario has been kept as general as possible, such back-reaction mechanism can be understood to be in effect in many varied situations of nonlinear nonequlibrium phenomenon. However, if we delve into specifics it is easy to derive a riogorous result for the noise correlations arising due to such backreacting terms in case of a quartic perturbation (which would be true for \textit{any even nonlinearity}).\\

\noindent The noise term for a quadratic Hamiltonian is given by  Eqn. (\ref{eq322}) and linear in the bath coordinates. Also, from observation we note that averages of the kind

\begin{eqnarray}
\langle \Pi(t)\Pi(t{'}) \rangle ^2 _{neq}
\label{eq328}
\end{eqnarray}

\noindent involve integrals of the form

\begin{eqnarray}
\int d\Omega F(q_{\mu},p_{\mu}) P ({q_{\mu}(0),p_{\mu}(0)})
\label{eq329}
\end{eqnarray}

\noindent with the usual interpretation of a phase space integral over the bath degrees of freedom. Notably, F is an even function in $p_{\mu}$ and $q_{\mu}$ or a bilinear combination of both. However, $P_{2}$ contains terms odd in bath coordinates as can be seen in Eqn.(\ref{eq327}). Hence, it can be concluded that upto first order in $\lambda t $, the noise correlations arising due to the Poisson Bracket induced evolution of a quenched distribution is zero.

\begin{eqnarray}
\langle \Pi(t)\Pi(t{'}) \rangle ^2 &=& 0
\label{eq330}
\end{eqnarray}

\noindent Hence, combining terms 1 and 2, for a quenched initial distribution modelled via a quartic nonlinearity, we find

\begin{eqnarray}
\langle \Pi(t)\Pi(t{'}) \rangle _{neq} &=& \langle \Pi(t)\Pi(t{'}) \rangle ^1 + \langle \Pi(t)\Pi(t{'}) \rangle ^2 \nonumber\\
&=& \sum_{\mu}^{N}\frac{\lambda^{2}}{\beta}\left[1+\epsilon(\frac{6a_{\mu}}{\beta m_{\mu}^{2}\omega_{\mu}^{4}}-1)\right]\frac{C_{\mu}^{2}}{m_{\mu}\omega_{\mu}^{2}}\cos{\omega_{\mu}(t-t^{'})}
\label{eq331}
\end{eqnarray}

\noindent The $\lambda^{2}$ piece is the usual FDR without the effects non-linearity present and it can be identified with  $\gamma(t-t')$ from Eqn.(\ref{eq302}). The second term of the order $\lambda^{2} \epsilon$ comes due evaluating $\Delta H_{1R}$ ie. $\langle \Gamma(t)\Gamma(t^{'})\Delta H_{1R} \rangle$. Also, notice that the whole term is obtained only by calculating a subset of correlations,that originally occur in the static case, appear in the non-equilibrium scenario. The reason is obvious: after the dynamics start, we must remember that the thermal environment is simply reduced to a quadratic bath. The excess correlations arising in the static case, due to the non-linearity of the noise for all times do not appear here. The classical correlation term that corrects the equilibrium case is a 3- point function involving the noise terms and the non-linear perturbation. Also, we should note that we use $L_{R}$ as the Louivillean in the first approximation of small $\lambda t $; therefore this extra term mentioned above contains all the information about the perturbation at $t<0$ in totality.\\

\noindent As previously stated by the authors, the convenient form of the non-equilibrium corrections led to the form-invariant redefinition of the FDR. Consequently, the damping term as well as the GLE can take a renormalised form as follows:
\begin{eqnarray}
\langle \Pi (t)\Pi (t^{'}) \rangle _{neq}&=& \gamma^{[R]}(t-t^{'})\\
M\ddot{X}&=& -V^{'}(X) -\int_{0}^{t}\gamma^{[R]}(t-t^{'})\dot{X}dt^{'} +\Pi(t)
\label{eq332}
\end{eqnarray}

\subsection {Some observations on the above model}

\noindent These two equations completely describe the effect of a small non-equilibrium perturbation on an open system. However, this also has major implications on a more fundamental level. A careful relook at the system-reservoir Hamiltonian immediately shows that the coupling between the system and bath co-ordinates are renormalised via the correction term that now scales with the temperature of the thermal ensemble at $t=0$. We have achieved what we proposed in the beginning: a non-equilibrium counterpart of the standard equilibrium GLE and FDR; however, now the couplings (at the Hamiltonian level) and the damping kernel (in the Langevin picture) are temperature dependent quantities and are also influenced by the nature of the non-linearities which dictate the perturbation.\\

\noindent Also, the strength of perturbation at the initiation of the dynamics on the ensemble of bath oscillators has been given no intrinsic structure. That is, though the $a_{\mu}$ -s are mode-dependent, an implicit distribution of how the modes are perturbed before the quench occurs is not established. This leaves space for some further exploration into the correlations above. Thus, we impose a distribution for the nonlinear modes whose degrees of freedom in effect are governed by the Canonical Ensemble at temperature T. Let,

\begin{eqnarray}
a_{\mu} &\propto& \omega _{\mu} ^ {\alpha} 
\label{eq333}
\end{eqnarray}

\noindent Depending on $\alpha$ , the character of the correlations differ significantly. Since in Eqn \ref{eq331}, the nonlinear correction term already contains an inverse power law in the bath mode frequencies  $\omega _{\mu}$. For $\alpha$ much lower than 6, the nonequilibrium correlations are dictated mostly by the low-wavelength modes as the high frequency contribution decreases fast. However, values of $\alpha$ close to 6 or more immediately reverses the trend; now high frequency modes dominate the correlations. Notably, a cut-off would be needed to regularise the nonlinearity parameter (like in the case of noise spectrum in the previous section) and also it should maintain the legality of the perturbative expansion. Another possibilty arises with the introduction of a random order in the nonlinearity parameter which we consider being drawn from a distribution with zero mean. A disorder-averaging of the noise correlations above then lead to the disappearance of any corrections for the nonequilibrium case. The model then mirrors that of an equilibrium scenario and this happens only because the correlations are linear in $a_{\mu}$.

\begin{eqnarray}
\langle \Pi(t)\Pi(t{'}) \rangle _{disorder} &=& \sum_{\mu}^{N}\frac{\lambda^{2}}{\beta} \frac{C_{\mu}^{2}}{m_{\mu}\omega_{\mu}^{2}}\cos{\omega_{\mu}(t-t^{'})}
\end{eqnarray}

\noindent Hence, a quenched initial condition of bath modes in the weak system-bath coupling limit (tractable for calculations) carry similar signatures of renormalised dissipation and temperature dependent extensions of noise correlations for relevant time scales of open system dynamics. The model is purely nonequilibrium in character driven by the initial jolt due to sudden change in bath mode structures.

\section{Summary}

\noindent The paper dealt with two important consequences of the nonlinearities introduced in the bath modes. Throughout, weak system-bath coupling scheme is assumed. On one hand, a detailed derivation of the emerguing bath spectrum from the nonlinear effective damping kernel is carried out. Markovian limits, correlations with memory and cut-off functions at high frequency limits are studied. In the quadratic bath case, it has been well established over the decades that an Ohmic approximation (linear bath spectrum) along with a frequency independent damping co-efficient leads to the Markovian dynamics of the system, a paradigm widely used in any stochastic process in all sciences. We show, in detail, that nonlinear baths require a Super-Ohmic piece to reach the Markovian limit and depends crucially on the strength and mass distribution of the bath modes. Likewise, conclusions about velocity correlations are determined from this information and different temporal scenarios are explored in the context of such noise correlations. The response of a probe to an external harmonic potential is included in this section. A second part to the paper lays down the framework for understanding quenched initial conditions and intrinsic nonequlibrium dynamics of a classical open system. The perturbation is modelled via a quartic nonlinearity and a Louivillean evolution equation is employed to calculate the time-dependent correlations arising out of such a picture.\\

\noindent Also, in studies pertaining to activated rate processes in classical as well as quantum domains, built up for almost a century based on the celebrated Arrhenius rate formula, it is worthwhile to expect that the correction terms deduced in this work may bring in new results. Nonlinear oscillators are also important in transport processes that describe heat conduction from one reservoir to another through a prescribed channel. In such processes where energy gets localized due to non-equipartition among the nonlinear modes, it is worthwhile looking into whether the correction terms can throw some new light on the various domains of diffusivities which come up in such contexts. In this regard the question of thermalisation, equilibration and presence of Nonequilibrium Stationary States can be investigated via numerical simulations of the open system dynamics obeying nonlinear coarse-grained version of the noise-correlations. Thus, a rigorous calculation of noise spectrum, dynamical correlations and time-dependent evolution of the bath kernels can serve as the basis for classical and quantum ventures into engineered reservoirs to fundamental questions in nonequilibrium physics.

\section{Appendix}

\subsection{Dissipation fucntion}
\begin{eqnarray}
\tilde {\gamma}(\omega) &=&  \frac{-i \omega}{M} \frac{2}{\pi} \int _{0}^{\infty} d\omega ' \frac{S(\omega ')}{\omega '} \frac {1}{\omega '^2 - \omega ^2 }\nonumber\\
&&=\lim _{p\to 0} \frac{-i \omega}{M} \frac{2}{\pi} \int _{0}^{\infty} d\omega ' \frac{S(\omega ')}{\omega '} \frac {1}{\omega '^2 - \omega ^2 +ipSgn(\omega)}\nonumber\\
&&=\lim _{p\to 0} \frac{-i \omega}{M} \frac{2}{\pi} \int _{0}^{\infty} d\omega ' \frac{S(\omega ')}{\omega '} \frac {1}{(\omega'-(\omega+ip))(\omega'+(\omega+ip))}
\end{eqnarray}

\noindent The poles are thus shifted above and below the real axis, at $\omega'=\omega+ip$ and $\omega'=-(\omega+ip)$. So we draw a contour as below.
\begin{figure}[h!]
 \includegraphics[width=\linewidth]{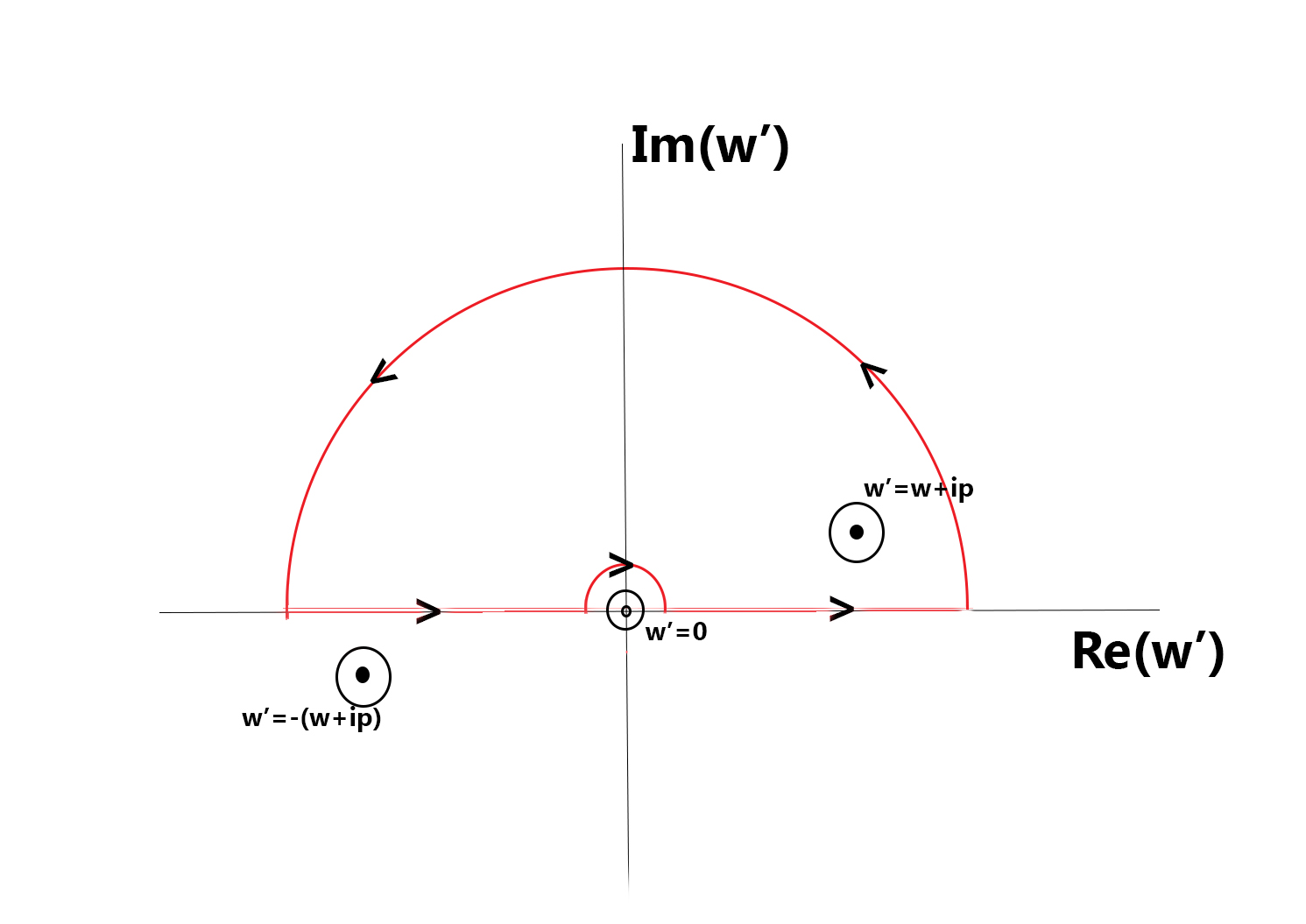}
\end{figure}\\

\noindent The arc integral vanishes since S($\omega$) must go to zero as $\omega$ goes to infinity. Also the spectral density must be zero at $\omega=0$.
Thus the real part of $\tilde{\gamma}(\omega$ is given by the imaginary part of the integral which by residue theorem is
$$lim _{p\to 0}\frac{1}{2}2\pi i\ \frac{S(\omega+ip)}{(\omega+ip)2(\omega+ip)}$$
$$=\frac{i\pi S(\omega)}{2\omega^2}$$
Putting this back, we obtain
$$Re[\tilde{\gamma}(\omega)]=\frac{S(\omega)}{M\omega}$$

\subsection{FDR in Fourier domain}

The noise correlations in time domain
\begin{eqnarray}
\langle\Gamma(t)\Gamma(t')\rangle=K_BT\sum\frac{C_\mu^2}{m_\mu\omega_\mu^2}cos\omega_\mu(t-t')=K_BTf(t-t')\nonumber
\end{eqnarray}
The noise correlations in frequency domain can be obtained by a double fourier transform as follows
\begin{eqnarray}
\langle\Gamma(\omega)\Gamma(\omega')\rangle=\frac{K_BT}{2\pi}\int_{-\infty}^{\infty}dt'e^{i\omega' t'}\int_{-\infty}^{\infty}dt e^{i\omega t}\langle\Gamma(t)\Gamma(t')\rangle\nonumber\\
=\frac{K_BT}{2\pi}\int_{-\infty}^{\infty}dt'e^{i t'(\omega+\omega')}\int_{-\infty}^{\infty}dt e^{i\omega (t-t')}\langle\Gamma(t)\Gamma(t')\rangle\nonumber\\
=\frac{K_BT}{2\pi}f(\omega)\delta(\omega+\omega')\nonumber
\end{eqnarray}

\begin{eqnarray}
f(\omega)=\int_{-\infty}^{\infty}f(t)e^{i\omega t}dt\nonumber\\
=\int_{-\infty}^{\infty}\sum\frac{C_{\mu}^2}{m_{\mu}\omega_{\mu}^2}cos\omega_{\mu}t(cos\omega t+isin(\omega t))dt\nonumber
\end{eqnarray}
The sine integral is zero because it renders the entire integrand to be an odd function. So 
$$f(\omega)=2Re[\int_{0}^{\infty}\sum\frac{C_{\mu}^2}{m_{\mu}\omega_{\mu}^2}cos(\omega_{\mu}t)e^{i\omega t dt}]$$
$$=2Re[\tilde{\gamma\omega}]$$
Thus,
$$\langle\Gamma(\omega)\Gamma(\omega'\rangle=\frac{K_BT}{\pi}Re[\tilde{\gamma(\omega)}]\delta(\omega+\omega')$$


\begin{thebibliography} {150}

\bibitem{chandra} S. Chandrasekhar, \emph{Stochastic Problems in Physics and Astronomy,
Reviews of Modern Physics}, \textbf{15}, No.1, 1-89 (1943)

\bibitem{lang} Don S. Lemons , \emph{Paul Langevin’s 1908 paper “On the Theory of Brownian Motion” [“Sur la théorie du mouvement brownien,” C. R. Acad. Sci. (Paris) 146, 530–533 (1908)]}, Am. J. Phys. \textbf{65} (11), 1079 (1997)

\bibitem{orn} G.E. Uhlenbeck and L.S.Ornstein, \emph{On the Theory of the Brownian Motion}
 Physical Review, \textbf{36},823 (1930) 

\bibitem{zwa1} R. Zwanzig, \emph{Nonlinear generalized Langevin equations}, Journal of Statistical Physics, \textbf{9}, 215 (1973).

\bibitem{mori} H. Mori, Transport, \emph{Collective Motion, and Brownian Motion}, Progress of Theoretical Physics  \textbf{33}, (3) 423 (1965).

\bibitem{zwa2} R. Zwanzig, \emph{Non-equilibrium Statistical Mechanics}, Oxford University Press (2001), USA

\bibitem{kawa} K. Kawasaki, \emph{Simple derivations of generalized linear and nonlinear Langevin equations}, Journ. of Phys. A: Mathematical and General, \textbf{6},9, 1289-1295 (1973)

\bibitem{weiss} U. Weiss, \emph{Quantum Dissipative System}, World Scientific (1999), Singapore.

\bibitem{feyn} R. P. Feynman and F. L. Vernon, \emph{The theory of a general quantum system interacting with a linear dissipative system},  Annals of Physics, \textbf{24}, 118 (1963).

\bibitem{heinz} Heinz-Peter Breuer and Francesco Petruccione, \emph{The Theory of Open Quantum Systems}, Oxford University Press (2007)

\bibitem{kubo} R. Kubo, \emph{The fluctuation-dissipation theorem}, Reports on Progress in Physics, \textbf{29}, 255 (1966).

\bibitem{vul2} U. M. B. Marconi, Andrea Puglisi, Lamberto Rondoni and Angelo Vulpiani, \emph{Fluctuation-Dissipation: Response Theory in Statistical Physics}, Physics Reports \textbf{461}, 111–195 (2008) 

\bibitem{risken} H. Risken , \emph{The Fokker-Planck Equation: Methods of Solution and Applications}, Springer Series in Synergetics (1996)

\bibitem{kamp} N.G. Van Kampen, \emph{Stochastic Processes in Physics and Chemistry}, North Holland (2007)

\bibitem{liu} Sha Liu, Peter Hänggi, Nianbei Li, Jie Ren, and Baowen Li, \emph{Anomalous Heat Diffusion},  Physical Review Letters, \textbf{112}, 040601 (2014)

\bibitem{bhadra1} C. Bhadra and D. Banerjee, Journ. of Stat. Mech., Vol 2016 (2016)

\bibitem{bhadra2} C. Bhadra, Journ. of Stat. Mech., Vol 2018 (2018)

\bibitem{jung} P. H\"angii and P. Jung, \emph{Coloured Noise in Dynamical Systems}, Advances in Chemical Physics, Vol. LXXXIX, John Wiley and Sons Inc.(1995)

\bibitem{bar} A. Barabasi and H. E. Stanley, \emph{Fractal Concepts in Surface Growth}, Cambridge University Press (1995)

\bibitem{leg1} A. O. Caldeira and A. J. Leggett, \emph{Path integral approach to quantum Brownian motion},  Physica, \textbf{121A}, 587-616 (1983).

\bibitem{dhar4} A Dhar, \emph{Heat Transport in low-dimensional systems}, Advances in Physics \textbf{57} (5), 457-537 (2008)

\bibitem{maes1} C. Maes,  \emph{On the second fluctuation--dissipation theorem for nonequilibrium baths}, Journal of Statistical Physics, \textbf{154}, 705 (2014).

\bibitem{maes2} C Maes and S Steffenoni, \emph{Friction and noise for a probe in a nonequilibrium fluid}, Physical Review E, 91 (2), 022128 (2015)

\bibitem{maes3} Urna Basu, Christian Maes and Karel Netočný, \emph{Statistical forces from close--to--equilibrium media}, New J. Phys., \textbf{17} 115006 (2015)

\bibitem{maes4} M Kruger and C Maes, \emph{The modified Langevin description for probes in a nonlinear medium}, Journal of Physics: Condensed Matter \textbf{29} (6), 064004 (2016)


















\end{thebibliography}
\end{document}